\documentclass[12pt]{article}
\usepackage[top=0.7in,bottom=1in,left=1.1in,right=1.1in,includehead]{geometry}
\usepackage[OT1]{fontenc}
\usepackage{epsfig,amsmath,amsfonts,verbatim,cite}
\usepackage{mathtext,marvosym,textcomp}
\usepackage{epsfig,amsmath,amsfonts,amssymb}
\usepackage{mathtools}
\usepackage{slashed}
\usepackage[usenames,dvipsnames,svgnames,table]{xcolor}
\usepackage{tikz}
\usetikzlibrary{arrows}
\usetikzlibrary{shapes.misc}
\usetikzlibrary{shapes,positioning,matrix,arrows}
\tikzstyle{every picture}+=[remember picture]
\tikzstyle{na} = [baseline=-.5ex]
\tikzstyle{format} = [rectangle,
                      thick,
                      minimum size=2cm,
                      draw=red!00!black!50,
                      top color=white,
                      bottom color=red!00!black!00,
                      text centered,
                      font=\fontsize{10}{10}\selectfont,
                      on grid]
\tikzstyle{format1} = [rectangle,
                      thick,
                      dashed,
                      minimum size=2cm,
                      draw=red!00!black!50,
                      top color=white,
                      bottom color=red!00!black!00,                     
                      text centered,
                      font=\fontsize{10}{10}\selectfont,
                      on grid]
\tikzstyle{format2} = [font=\fontsize{10}{10}\selectfont,
                      on grid]
\tikzset{cross/.style={cross out, draw=black, minimum size=2*(#1-\pgflinewidth), inner sep=0pt, outer sep=0pt},
cross/.default={5pt}}
 \usepackage[ normalem]{ulem}
 \usepackage{tocloft}

 \usepackage[debug,pageanchor=false]{hyperref}
\hypersetup{colorlinks=true,linktocpage,breaklinks,
            urlcolor=blue,
            linkcolor=red,
            citecolor=blue,
            }

\def\a{\alpha} \def\b{\beta}  \def\d{\delta} 
  \def\h{\eta} \def\q{\theta}
    \def\m{\mu}
\def\n{\nu} \def\x{\xi}   \def\r{\rho}
 \def\s{\sigma}   \def\f{\varphi}

 \def\S{\Sigma}  
\def\F{\Phi}   \def\W{\Omega}

\def\fr{\frac}  \def\dt{\partial}

\def\ph{\phantom}
\def\mc{\mathcal}

\def\mH{\mathcal{H}}

\def\tX{\tilde{X}}

\def\tY{\tilde{Y}}

\def\tx{\tilde{x}}

\def\tz{\tilde{z}}

\def\tk{\tilde{k}}

\def\tm{\times}
\def\bt{\bullet}

\def\hdt{\hat{\dt}}

\def\XX{\mathbb{X}}

\newcommand\bqa {\begin{eqnarray}}
\newcommand\eqa {\end{eqnarray}}

\newcommand{\bear}{\begin{array}}
\newcommand{\enar}{\end{array}}

\newcommand{\be}{\begin{equation}}
\newcommand{\ee}{\end{equation}}
\def\bea{\begin{eqnarray}}
\def\eea{\end{eqnarray}}

\begin{document}
\renewcommand{\contentsname}{}
\renewcommand{\refname}{\begin{center}References\end{center}}
\renewcommand{\abstractname}{\begin{center}\footnotesize{\bf Abstract}\end{center}} 
 \renewcommand{\cftdot}{}

\begin{titlepage}

\ph{1}

\vfill

\begin{center}
   \baselineskip=16pt
   {\Large \bf  T-duality and entanglement entropy\\for NS five-branes. }
   \vskip 1cm
    Anastasia A. Golubtsova$^{1,2}$, Edvard T. Musaev$^{3,4}$
       \vskip .6cm
             \begin{small}
                          {\it $^1$Bogoliubov Laboratory of Theoretical Physics, JINR,                       141980 Dubna, Moscow region, Russia\\
                          $^2$Dubna State University,
                          Universitetskaya str. 19, Dubna, 141980, Russia} \\
                          {\tt golubtsova@theor.jinr.ru}\\[0.2cm]
                          {\it $^3$Moscow Institute of Physics and Technology,
                          Institutskii per. 9, Dolgoprudny, 141700,  Russia, \\
                          $^4$Kazan Federal University, Institute of Physics,
                          Kremlevskaya st. 16, Kazan, 420008, Russa} \\
                          {\tt musaev.et@phystech.edu} 
\end{small}
\end{center}

\vfill 
\begin{center} 
\textbf{Abstract}
\end{center} 
\begin{quote}
In this work an algorithm is proposed to calculate entropy for field theories living on NS five branes, which gives a result invariant under T-duality. This is a deformation of the well known Ryu-Takayanagi formula, which takes into account dependence of localized backgrounds for the branes on winding modes of strings.
\end{quote} 
\vfill
\setcounter{footnote}{0}
\end{titlepage}

\setcounter{page}{2}
\tableofcontents

\section{Introduction}

Entanglement entropy is a non-local observable which measures entanglement between two subsystems of a quantum system.
It has many applications in studies of phenomena in quantum gravity, quantum information, condensed matter and high energy physics.
Particularly, entanglement entropy in the context of the gauge/gravity duality is aimed to shed some light on understanding of quantum gravity into the bulk \cite{Rangamani:2016dms}.

Following the holographic prescription, the  entanglement entropy between a subsystem (region) $A \in\mathbb{R}^{d}$ that has a  $d-1$-dimensional boundary $\partial A$ and a remaining part $B$  can be calculated by Ryu-Takayanagi formula \cite{Ryu:2006bv,Ryu:2006ef,Hubeny:2007xt}
\begin{eqnarray}\label{1.1}
S =  \frac{\text{Area}(\gamma_{A})}{4G^{d+2}_{N}},
\end{eqnarray}
where $\gamma_{A}$ is the minimal $d$-dimensional surface in $AdS_{d+2}$ space whose boundary coincides with the boundary of the region $A$ ($\partial A = \partial \gamma_{A}$), $G^{d+2}_{N}$ is $d +2$-dimensional Newton constant.

For the classic  case with $AdS$ on the gravity side, which geometry is not supported by any scalar field,
 and conformal theory of QFT side the area of the surface $\gamma_{A}$ is defined through the induced metric  by the relation
\begin{equation}\label{1.1a}
A = \int d^{d}\sigma \sqrt{|\det{G_{\alpha\beta}}|},
\end{equation}
where  $G_{\alpha\beta}=g_{MN}\partial_{\alpha}X^{M}\partial_{\beta}X^{N}$ is the induced metric of $\gamma_{A}$ and $g_{MN}$ is the metric of the background. Important examples of $AdS$ spacetimes include near-horizon geometries of $p$-branes.
In the form (\ref{1.1a}) it can be applied to studies of entanglement entropy for non-dilatonic branes, namely D3, M2 and   M5 branes \cite{Quijada:2017zif}.

The  generalization of the entangled functional (\ref{1.1}) with (\ref{1.1a}) for branes with non-conformal boundaries reads
\bea\label{RTncbr}
S = \int d^{8}\sigma \frac{1}{4G^{10}_{N}} \sqrt{|\det G_{ind}|}e^{-2\phi},
\eea
where $\phi$ is the dilaton.
The holographic entanglement entropy for configurations on D2 and NS five-branes was calculated in the original work  \cite{Ryu:2006ef},
on D3 and  D4 branes in \cite{Klebanov:2007km,Pakman:2008ui,Arean:2008az}, on D1-D5 brane intersection in \cite{Asplund:2011cq}.
 The dilaton destroys the scale symmetry, 
but  we still can detect  a certain field theory on the boundaries of the branes and discuss a holographic picture.
For example, 
for NS5 brane in the long distances of the theory is governed by  the (2, 0) SCFT for IIA theory and the IR free SYM with sixteen supercharges for IIB, while
the short distance behavior leads to a linear dilaton geometry   \cite{Aharony:1998ub} that can be described through the so called Little String Theory  $\mc{N}=(2,0)$ and $\mc{N}=(1,1)$ on the Type IIA and Type IIB NS5 branes respectively.

In this work we aim at studying T-duality aspects of entanglement entropy for field theories living on NS five branes, including the exotic branes $5_2^r$ with $r=0,1,2,3,4$. For the NS5 brane the decoupling limit is known to be LST, which is a 6-dimensional theory describing dynamics of string-like degrees of freedom which do not have gravitational modes in their spectrum. In all other respects they exhibit essentially stringy behaviour, such as Hagedorn temperature and T-duality of spectrum \cite{Losev:1997hx, Kutasov:2001uf, Aharony:1999ks}. This is due to the fact that in contrast to D-branes the decoupling limit for the NS branes does not involve taking $\a'\to0$. This preserves stringy properties on the world-volume. The origin of T-duality in LST is the simple observation that a compactified NS5 brane transform into itself under T-duality along a world-volume direction. One the language of LST this transform into T-duality symmetry of the 6d theory with one compact direction, the direct analogue of that of the 10d string theory.

In addition however one may wonder what are the properties of the theory under T-duality transformations in the transverse directions, i.e. those which change the brane, say from NS5 brane to the KK5-monopole. By simple counting of degrees of freedom one concludes that the theory does not change under that. I.e. the Type IIA/B NS5 brane carries the same world-volume theory as the Type IIB/A Kaluza-Klein monopole \cite{Sen:1997js}. Continuing this logic one concludes that the theory should not change along the whole T-duality orbit.
\begin{equation}
\begin{aligned}
5_2^0(A/B) && \longleftrightarrow && 5_2^1(B/A)\longleftrightarrow && 5_2^2(A/B)\longleftrightarrow && 5_2^3(B/A)\longleftrightarrow && 5_2^4(A/B)
\end{aligned}
\end{equation}
Here we use the notations for the branes of \cite{Obers:1998fb} (see also \cite{deBoer:2012ma} for more on that), and the last three are exotic. The fact that the corresponding world-volume field theories do not change under T-duality trivially follows from the T-duality invariant world-volume effective action for these branes presented in \cite{Blair:2017hhy}. This is a single action for the whole orbit, which drops into actions for a representative upon removing half of the scalar fields living on the brane (geometric or dual coordinates). Since from the world-volume point of view these are just scalar fields moving in a dynamical background, replacement one by its dual does not change anything for it.

However, applying the Ryu-Takayanagi prescription for geometric entropy to the background of say Kaluza-Klein monopole one gets the answer which is different from the one for the NS5 brane, which clearly breaks the T-duality invariance. In this paper we show that the reason for that is that in its geometric and straightforward form this prescription does not take into account dependence on the winding direction of the localized Kaluza-Klein monopole. Indeed, in \cite{Tong:2002rq, Harvey:2005ab,Kimura:2013fda,Kimura:2018hph} it has been shown that instanton corrections coming from the 2d sigma-model describing the KK5 background, change the geometry such that field start depending on a winding mode. This correct the throat behaviour of the KK5-monopole to make it the same as that of the NS5 brane. In \cite{Jensen:2011jna, Berman:2014jsa, Bakhmatov:2016kfn}  it has been shown that this has simple explanation in terms of Double Field Theory, that is to do a T-duality transformation in a direction $z$ one replaces $z$ by it dual $\tilde{z}$ in all expressions. The same is true for producing exotic backgrounds, and the corresponding instanton interpretation has been presented in \cite{Kimura:2013zva}. In this work we consider the invariant action of \cite{Blair:2017hhy} and propose an algorithm to calculate entanglement entropy for theories living on branes with non-trivial dynamics in doubled space.

This paper is structures as follows. In Section \ref{geom} we present a short technical review of how the geometric entanglement entropy is calculated and explicitly show that the RT formula gives different results when applying to NS five-brane backgrounds belonging to the same T-duality orbit. In Section \ref{dft} we turn to invariant dynamics governed by the action of \cite{Blair:2017hhy}, shortly review how one obtains different action from the invariant one, and describe the algorithm which produces an invariant answer for entanglement entropy. In addition we comment on the geometric meaning of the expression, which is an important and subtle point due to lack of the notions of integration, distance and area in doubled geometry.

\section{Geometric entanglement entropy}
\label{geom}

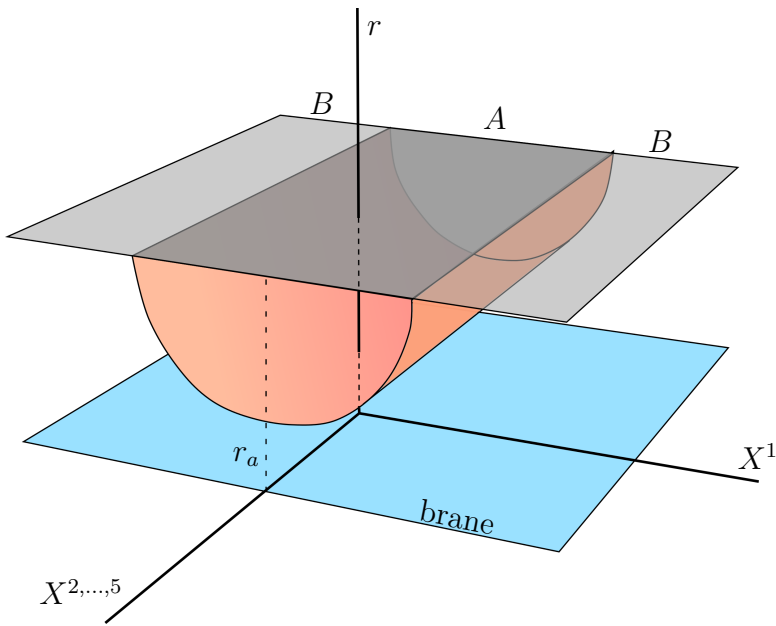
\begin{figure}[ht]
\centering
\includegraphics[width=10cm]{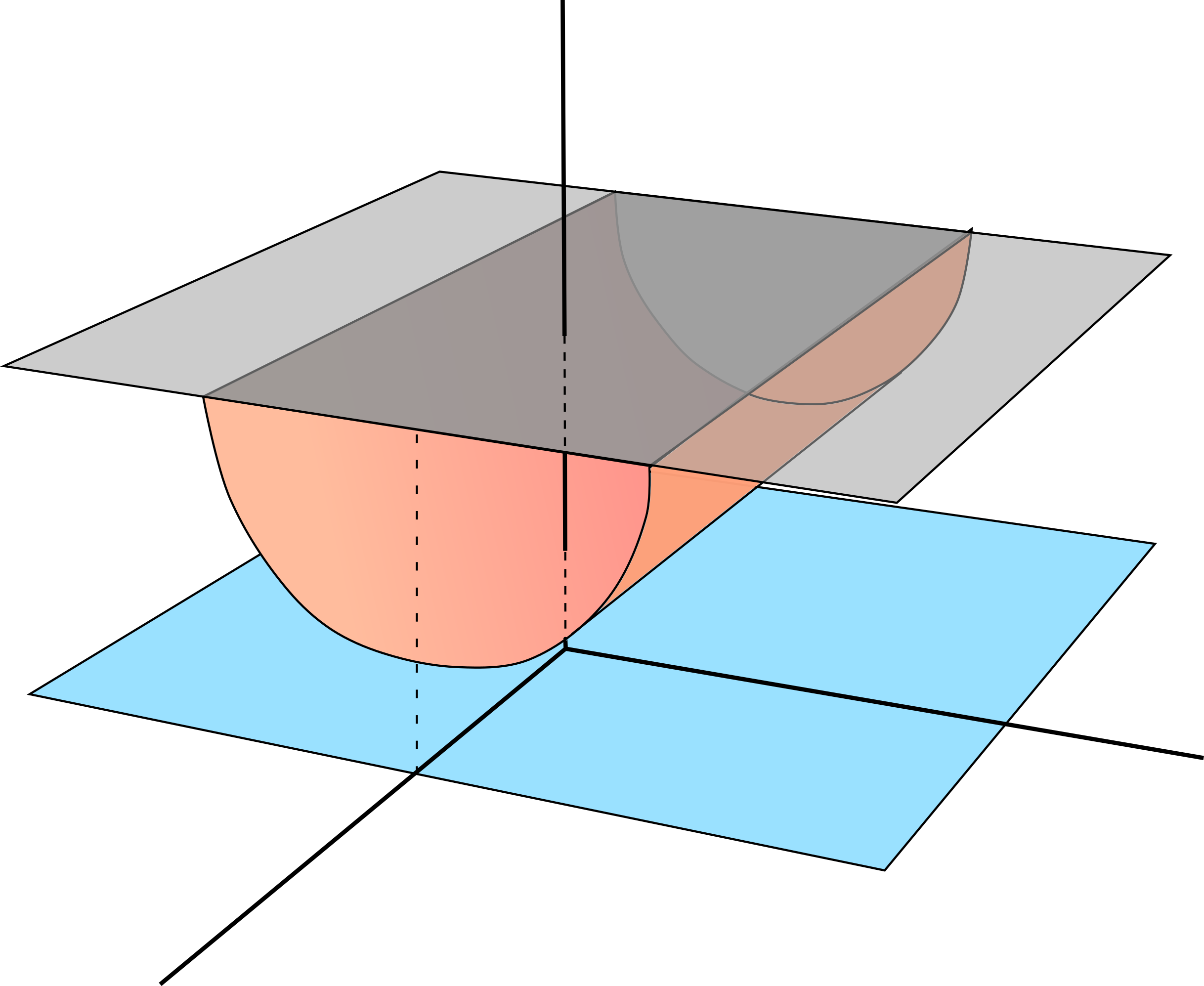}
\caption{Configuration of the embedding}
\label{embed}
\begin{tikzpicture}[overlay]
\node at (-0.1,9.5) (r) {$r$};
\node at (-1.8,3.8) (ra) {$r_a$};
\node at (-4,2) (X25) {$X^{2,\dots,5}$};
\node at (5,3.8) (X1) {$X^1$};
\node at (1.5,8.3)  (A) {$A$};
\node at (3.7,8)  (B1) {$B$};
\node at (-0.8,8.5)  (B2) {$B$};
\node[rotate=-11] at (1,3) (br) {brane};
\end{tikzpicture}
\end{figure}

The usual choice of areas which carry entangled states, which significantly simplifies calculations, is the infinite strip set-up. For that one considers a surface in the space transverse to a brane one which the field theory lives (shaded on Fig. \ref{embed}). This surface is the boundary for the minimal surface, which tends to curve closer to the brane due to the transverse geometry. For D-branes this surface is identified with the AdS conformal boundary. The geometric formula of Ryu and Takayanagi gives entanglement entropy of states in the region A and B on the picture.

In this section the standard formula for calculation of geometric entanglement entropy is applied to the standard NS5 brane and to the KK5-monopole and exotic $5_2^2$ brane. Due to the special circle already for the KK5 background one gets expressions very different from that for the NS5 background. From this we conclude that one must develop a different algorithm and a different understanding of the RT expression to properly capture transformations along the NS five-brane T-duality orbit.

\subsection{Non-conformal theories: geometric five-branes}
\label{direct_KK}

When turning to NS five branes one encounters 6d theories which describe string-like degrees of freedom which do not have gravitational excitation in their spectrum, the so-called Little String Theory. In the field theory limit these drop to non-conformal field theories since the corresponding brane backgrounds contain non-trivial dilaton and are not asymptotically AdS. However, for these one also can define entanglement entropy using the Ruy-Takayanagi conjecture (\ref{RTncbr}) and write
\begin{equation}
S=\int_{\S} d^5 \s e^{-2\f} \sqrt{\det G_{\a\b}},
\end{equation}
where $\{\s^\a\}$ wuth $\a=1,\dots,5$ are coordinates on the space-like surface $\S$  and $G_{\a\b}=\dt_\a X^\m \dt_\b X^\n G_{\m\n}$ is the induced metric on the surface. For our purposes we choose the simplest shape for the surface $\S$  generated by an infinite stripe. Theory for which the entanglement entropy is calculated lives on a surface parallel to the NS5 brane placed at some $r=r_b$. Entanglement is assumed for the states defined in the interior $A$ and exterior $B$ regions of the grey surface on the Fig.\ref{embed}. According to the conjecture this is equal to area of the minimal surface $\S$ whose boundary satisfies $\dt \S=\dt A$.

Embedding of the brane and of the surface is given by
\begin{equation}
\begin{aligned}
&      &&  0 && 1 && 2 && 3 && 4 && 5 && r && \q_1 && \q_2 && \f \\
\hline
& NS5  &&\tm &&\tm&&\tm&&\tm&&\tm&&\tm&&  \\
& \S   &&    &&\bt&& L && L && L && L && \bt
\end{aligned}
\end{equation}
where $\tm$ denote the world-volume directions. The surface $\S$ extends from $-L$ to $L$ in the directions denoted by $L$ above, while it is somehow curved in the directions denoted by bullets. I.e. one can choose coordinates and embedding functions for the surface as follows
\begin{equation}
\begin{aligned}
X^1&=X(r),\\
X^{2,\dots,5}&=\s^{2,\dots,5},\\
r&=\s^1.
\end{aligned}
\end{equation}

Background for the NS5 brane is given by
\begin{equation}
\begin{aligned}
ds^2&=\h_{rs}dx^rdx^s+H(dr^2+r^2\d\W_{3}^2),\\
\mH&=dB,\\
e^{-2(\f-\f_0)}&=H(r)^{-1},
\end{aligned}
\end{equation}
with the harmonic function $H(r)=1+h/r^2$. Hence one writes for the entropy
\begin{equation}
\begin{aligned}\label{RT-ns5}
S_{NS5}=
16L^4 \int dr H(r)^{-1}\sqrt{H(r)+X'(r)^2}.
\end{aligned}
\end{equation}
The usual minimisation procedure implies that the embedding function $X(r)$ should satisfy
\begin{equation}
\label{Xemb}
X_{NS5}'(r)=\pm\fr{H(r_a)^{1/2}H(r)}{\sqrt{H(r_a)^2-H(r)^2}},
\end{equation}
where we used the condition that $X'(r_a)=0$, which basically means that $r_a$ is the turning point for the surface $\S$. Note, that one has to set $r>r_a$ to keep the expression in the square root positive, which means that the turning point is closer to the brane than the surface $r=r_b$ on which the  field theory is defined. This is the usual configuration for the AdS/CFT correspondence and hence the initial setup and the Fig.\ref{embed}.

One can apply the same procedure to the worldvolume theory of the KK-monopole which for the Type IIA(B) monopole is the same as for the Type IIB(A) NS5 brane. Background geometry is given by the following configuration
\begin{equation}
\begin{aligned}
ds^2&=\h_{rs}dx^rdx^s+H^{-1}(d\tz+A_idy^i)^2+H\d_{ij}dy^i dy^j,\\
B&=0,\\
e^{-2(\f-\f_0)}&=1.
\end{aligned}
\end{equation}
Here $\tz$ is the normal geometric coordinate used to measure distances in space-time, however it is dual to the coordinate $z$ of the corresponding NS5 background. Note that the harmonic function is smeared $H=1+h/r$.

Repeating the same calculation as above one gets for the entropy and for the embedding function $X_{KK5}(r)$
\begin{equation}
\begin{aligned}
S_{KK5}&=16L^4 \int dr \sqrt{H(r)+X'(r)^2},\\
X_{KK5}'(r)&=\pm C\sqrt{H(r)}, \quad C=\mbox{const}.
\end{aligned}
\end{equation}
One first notices that the crucial difference with the previous case, that is $dr/dX=0$ at $r=0$, i.e. on the brane itself, while for the NS5 brane background the turning point is at some $r_a\neq 0$. This can be understood in terms of the short distance behaviour of NS5 branes and KK5 monopole. As it has been shown in \cite{Tong:2002rq,Gauntlett:1992nn} the former is the  version of the H-monopole (which is the proper T-dual of the KK monopole) localized due to instanton corrections. However, the localization breaks isometry along the compact circle of H-monopole and one observes a throat behaviour at short distances. 

To cure the near-brane behaviour of the KK5-monopole background one also considers instanton corrections \cite{Harvey:2005ab}. Only in this case one may expect result for the entropy which reproduce those for the NS5 brane. Such corrections however deform the background by introducing a non-trivial dependence on string winding coordinates, which requires double field theory to consistently address the issue, as in \cite{Jensen:2011jna,Bakhmatov:2016kfn}.

\subsection{Exotic five-branes}
\label{direct_exotic}

Hence, the answer for the entropy which one obtains for the theory living on the KK monopole is different from that for the NS5 brane. The important point here is that although T-duality exchanges IIA and IIB branes the theories living on the NS5A(B) and the KK5B(A) are the same and the entropy should not change. When going further along the T-duality orbit towards exotic branes the situation does not get better. Smearing the KK5 background along $y_3$ and T-dualizing one arrives at the exotic $5_2^2$-brane with background given by \cite{deBoer:2012ma}
\bea
ds^{2} &=& \eta_{rs}dx^{r}dx^{s} + HK^{-1}\left(d\tilde{z}^{2} + d\tilde{y}^{2}_{3}\right) + H \delta_{\alpha\beta}dy^{\alpha}dy^{\beta}, \\
B& = & h\theta K^{-1} d\tilde{z}\wedge d\tilde{y}_{3}, \\
e^{-2(\phi - \phi_{0})} &=& HK^{-1},\\
K &= & H^{2} + (h\theta)^2.
\eea
Here the harmonic function is further smeared $H(r)=1+h \log r$ and does not behave well at space infinity

This background is globally well-defined only up to a monodromy around the brane, hence the non-geometric properties of the background. Naively applying the above procedure one obtains
\begin{equation}
\begin{aligned}
S_{5_2^2}&=16L^4\int d r \fr{H(r)\sqrt{H(r)+X'(r)^2}}{H(r)^2+h^2 \q^2},\\
X_{5_2^2}'&=\pm\fr{C \sqrt{H(r)}(H(r)^2+h^2 \q^2)}{\sqrt{H(r)^2-C^2 (H(r)^2+h^2 \q^2)^2}}.
\end{aligned}
\end{equation} 
The embedding function $X_{5_2^2}(r)$ delivering extremum to $S_{5_2^2}$ is apparently not well-defined and moreover it explicitly depends on $\q$. Hence, the entropy also depends explicitly on the coordinate $\q$ respecting the monodromy property of the background.

On the other hand, the worldvolume theory on $5_2^2$-brane should not differ from that of the KK-monopole or NS5 brane (with proper replacement of Type IIA with Type IIB). To perform calculation of entanglement entropy for such theories which respect T-duality we use the T-duality covariant action of \cite{Blair:2017hhy} for the 5-brane orbit. It suggests that the worldvolume theory is the same irrespective of the choice of the brane (equivalently, the section constraint or orientation in the doubled space) upon the proper exchange of the worldvolume scalars $X^\m$ with their duals $\tX_{\m}$.

\section{Entanglement entropy in  DFT}
\label{dft}

Double Field Theory being a T-duality covariant formulation of supergavity (string theory) allows to consider the whole T-duality orbit instead of a single representative. In this section we propose a deformation of the geometric prescription for entanglement entropy and embed the expression for entropy itself into the DFT framework. Let us start with brief description of how NS five-branes are embedded into doubled space.

\subsection{Embedding of NS five-branes in doubled space}

In \cite{Blair:2017hhy} it was shown in details how one can construct a T-duality covariant action for NS five-branes. The covariancy here is understood in the following way: one has a single expression which is written in terms of DFT (covariant) fields and which reproduces the effective action for the NS5B-brane, KK5A monopole and exotic branes $5_2^2$B, $5_2^3$A, $5_2^4$B. The full action smartly chooses these frames depending on which symmetries of the doubled spaces are eventually realized on the world-volume. Let us briefly describe the process focusing only on the NS-NS sector and only on the DBI part of the action, which is given by
\begin{equation}
\label{full_5}
S_{NS,DBI}[Y(\x)]=\int_V d^6 \x e^{-2d}\sqrt{\det h_{ab}}\sqrt{\displaystyle -\det\Big(g_{\m\n}\dt_\a{X^\m} \dt_\b X^\n+ \mH_{MN}\hat{D}_\a Y^M \hat{D}_\b Y^N\Big)},
\end{equation}
where we introduce
\begin{equation}
\begin{aligned}
h_{ab}&=k_a^M k_b^N \mH_{MN},\\
\hat{D}_\a Y^M&=\hdt_\a Y^M+\dt_\a X^\m A_\m{}^M,\\
\hdt_\a Y^M&=\dt_\a Y^M-(h^{-1})^{ab}k_a^M k_b^N\mH_{NP}\dt_\a Y^P,\\
\mH_{MN}&=
\begin{bmatrix}
G_{mn}-B_{m}{}^k B_{kn} & B_{n}{}^q\\
B_m{}^p & G^{pq}
\end{bmatrix}.
\end{aligned}
\end{equation}
Here the full space-time is split into the part parallel to the five-brane, labelled by the indices $\m,\n=\{0,5\}$, and the part transverse to the branes, which is doubled and labelled by $M,N,P,Q=\{6,7,8,9,\tilde{6},\tilde{7},\tilde{8},\tilde{9}\}$. The vector fields $A_\m{}^M$ result from the Kaluza-Klein decomposition of the full 10D theory 
\begin{equation}
A_\m{}^M=
\begin{bmatrix}
A_\m{}^m \\
-B_{\m m}
\end{bmatrix}.
\end{equation}
The integration is performed over world-volume of the brane which is parametrized by six coordinates $\{\x^\a\}$. The hatted derivative $\hdt_\a$  contains a projector part and is designed in such a way as to always remove half of the fields $Y^M$ from the action. Upon adding the action for DFT fields this results in field configurations which do not depend on half of DFT coordinates and hence is a worldvolume realization of the section constraint. Finally, the choice of the section frame and hence a representative of the T-duality orbit is done by choosing the particular form of the vectors $k_a{}^M$, which must satisfy the following algebraic section constrain
\begin{equation}
k_a^Mk_b^N\h_{MN}=0,
\end{equation}
where $\h_{MN}$ is the usual O(4,4) invariant tensor 
\begin{equation}
\h_{MN}=\begin{bmatrix}
0 & 1 \\
1 & 0
\end{bmatrix}
\end{equation}
and the indices $a,b={1,4}$ enumerate the Killing vectors. The reason why we call these vectors Killing will be clear in a moment.

For the O(4,4) configuration there exist five inequivalent solutions of the algebraic section constrain, each of which corresponds to the branes $5_2^r$ with $r=0,1,2,3,4$ showing the number of quadratic direction in the mass of the corresponding 3D BPS state (see \cite{deBoer:2012ma} for more detailed description of these notations). Here we list five representative solutions
\begin{equation}
\label{Kill}
\begin{aligned}
NS5=5_2^0: && k_a^M&=(0,0,0,0;\tk_{a 1}, \tk_{a 2},\tk_{a 3},\tk_{a 4}),\\
KK5=5_2^1: && k_a^M&=(0,0,0,k_{a}^{4}; \tk_{a 1},\tk_{a 2},\tk_{a 3}, 0),\\
Q=5_2^2: && k_a^M&=(0,0,k_{a}^{3},k_{a}^{4};\tk_{a 1},\tk_{a 2},0,0),\\
R=5_2^3: && k_a^M&=(0,k_{a}^{2},k_{a}^{3},k_{a}^{4};\tk_{a 1},0,0,0),\\
R'=5_2^4: && k_a^M&=(k_{a}^1,k_{a}^{2},k_{a}^{3},k_{a}^{4};0,0,0,0).
\end{aligned}
\end{equation}
For example, for the NS5 brane case, which is the first line above, one chooses all vectors $k_a^M$ to be along the dual coordinates. Substituting this back into the action one checks that all fields $Y_m$ drop from the expression rendering field configurations independent on the corresponding DFT coordinates. This is due to
\begin{equation}
\hdt_\a Y_m=B_{mn}\dt_\a Y^n,
\end{equation}
where $B_{mn}$ is the usual Kalb-Ramond two-form gauge field. The same is true for all other configurations up to the R'-brane which is a co-dimension-0 object from the point of view of the conventional supergravity.

To obtain explicit expression for the background fields for a fixed choice of the Killing vectors, one considers the full action with the embedding given by Dirac delta functions $\d^{(8)}(\XX^M-Y^M(\x))$, where $\XX^M=(x^m,\tx_m)$ are the coordinates of DFT. The reparametrization invariance of the world-volume is fixed as usual as
\begin{equation}
\label{gauge_fix0}
\begin{aligned}
X^\a&=\x^\a.
\end{aligned}
\end{equation}
Consider for example the KK-monopole, which is the second line above, where the fields $Y_{1,2,3}$ and $Y^4$ drop from the action meaning that the field configurations as functions are of the form $H=H(x^1,x^2,x^3,\tx_4)$. This is interpreted as a functional dependence of the background on three geometric coordinates $x^{1,2,3}$ and one non-geometric (dual or winding) coordinate $\tx_4$. This is due to an additional piece of information fixed in the DFT action, where one always understands $\XX^m=x^m$ as geometric coordinates, i.e. those used to measure space distances, and $\XX_m=\tx_m$ as their non-geometric duals. Without this fixing one will just count each brane four more times obtaining the same backgrounds but with different names for the same physical coordinates.

Such dependence of exotic backgrounds (starting from the KK monopole) on dual coordinates has been shown for the DFT monopole in \cite{Bakhmatov:2016kfn} and will be important for our further discussion.

\subsection{Invariant entropy and minimal surface}

The main feature of the effective action \eqref{full_5} is that it does not depend on the choice of the T-duality frame and describes dynamics of all five-branes dual to NS5 brane. Since the world-volume theory does not change when switching from (Type IIB) NS5 brane to (Type IIA) KK5-monopole, the corresponding entanglement entropy should not change as well. One can conjectures the following deformation of the Ryu-Takayanagi formula which provides such invariant description:
\begin{equation}
\label{inv_entr}
S_5=\int_\S d^5 \s e^{-2d}\sqrt{\det h_{ab}}\sqrt{\displaystyle -\det\Big(g_{\m\n}\dt_\a{X^\m} \dt_\b X^\n+ \mH_{MN}\hat{D}_\a Y^M \hat{D}_\b Y^N\Big)},
\end{equation}
where the notations are the same as before. In a moment we will explicitly show that this expression gives the usual RT formula whose minimization gives the geometric entanglement entropy for the NS5 brane case. For other representatives of the orbit one gets a deformation of the formula, however the integral itself does not distinguish between the allowed choices of the duality frame.

Before that it is important to discuss the meaning of the integration and of the surface $\S$ here. Going back to the effective action \eqref{full_5} one notes that the integration there is performed over the world-volume $V$ parametrized by $\s^\a$, which is a usual geometric manifold with properly defined integration measure. On this manifold one defines $6+(4+4)$ fields $\{X^\m(\x),Y^M(\x)\}$, which are identified with coordinates in the space-time and the doubled coordinates of the O(4,4) DFT. The crucial point here is that without such identification, these fields do not carry the meaning of coordinates on a doubled space and hence one is not actually doing doubled geometry, and rather works with a number of fields. For more discussion on this see \cite{Blair:2017hhy}.

Although the expressions \eqref{full_5} and \eqref{inv_entr} look almost the same, there is fundamental difference between them. While in action one varies with respect to the background fields keeping the embedding fixes, for the entropy the background is fixed by our choice of the brane and variation goes with respect to the embedding. The latter is defined by identification of the surface coordinates $\{\s^\a\}$ with the fields $X^\m,Y^M$, which define the (doubled) space-time dependence of the background. Since, the harmonic function depends only on a singlet combination $r$, a natural choice of the embedding is
\begin{equation}
\label{gauge_fix}
\begin{aligned}
X^{2,3,4,5}&=\s^{2,3,4,5},\\
\s^1&=r,
\end{aligned}
\end{equation}
and the remaining field $X^1=X(\s^1)$ is a function delivering minimum to the expression. Here the particular form of the field $r$ depends on the choice of the background and reads
\begin{equation}
\begin{aligned}
NS5=5_2^0: &&  r^2&=(Y^1)^2+(Y^2)^2+(Y^3)^2+(Y^4)^2,\\
KK5=5_2^1: && r^2&=(Y^1)^2+(Y^2)^2+(Y^3)^2+(\tY_4)^2,\\
Q=5_2^2: && r^2&=(Y^1)^2+(Y^2)^2+(\tY_3)^2+(\tY_4)^2,\\
R=5_2^3: && r^2&=(Y^1)^2+(\tY_2)^2+(\tY_3)^2+(\tY_4)^2,\\
R'=5_2^4: && r^2&=(\tY_1)^2+(\tY_2)^2+(\tY_3)^2+(\tY_4)^2.
\end{aligned}
\end{equation}
These follow from solutions of the equations of motion for the full action $S_{DFT}+S_{brane}$ which boil down to Poisson equation with delta source whose solution is the harmonic function $H=H(r)$ with $r$ given by the above expression. The number of dual coordinates entering the dependence of the fields is equal to the number of special circles.

The gauge fixing conditions  \eqref{gauge_fix} can be understood as a proper embedding of the surface $\S$ in the doubled 5+(4+4)-dimensional space. This is similar to the way how the magnetic charge for these branes has been calculated in \cite{Bakhmatov:2016kfn}, however now the integration remains proper integration over a conventional manifold with conventional measure. The structure of the doubled space shows up only at the level of the Killing vectors and of the interaction between the effective action and the full DFT action. Before that, the integration does not distinguish between $Y^m$ and $\tY_m$, as it should be since the corresponding world-volume theories do not feel this as well. The integration is then performed in $\s^{2,3,4,5}\in [-L,L]$ for some large $L$ and from the points $X'=0$ in the $\s^1$ direction. This is what is usually called the rectangular strip area, which is the simplest to perform calculations. In principle, one may choose a different embedding which will correspond to a different area inside the world-volume theory.

Let us postpone the discussion, of how this process is seen from the point of view of the world-volume theory, to the Discussion section and now move to explicit examples to show invariance of the expression.

\subsection{Explicit examples}

Let us start with the T-duality frame which corresponds to NS5 brane, which fixes the Killing vectors to be
\begin{equation}
k_a^M=(0;\tk_{am}).
\end{equation}
Then the matrix $h_{ab}$ becomes $h_{ab}=\tk_{am}\tk_{bn}g^{mn}$ and one has 
\begin{equation}
\det{h_{ab}}=|\tk|^2 g^{-1},
\end{equation}
where $|\tk|=\det\tk_{am}$ and $g=\det G_{mn}$. The inverse of the matrix $h_{ab}$ is then
\begin{equation}
(h^{-1})^{ab}=(\tk^{-1})^{am}(\tk^{-1})^{bn}G_{mn},
\end{equation}
where $(\tk^{-1})$ is the inverse of $\tk_{am}$ understood simply as a $4\times 4$ matrix. Hence, for derivatives of the fields $Y^M$ we have
\begin{equation}
\begin{aligned}
\hdt_\a Y^m&=\dt_\a Y^m,\\
\hdt_\a \tY_m&=\dt \tY_m-(h^{-1})^{ab}\tk_{am}\tk_{an}\mH^n{}_P\dt_\a Y^p=B_{mn}\dt_\a Y^n.
\end{aligned}
\end{equation}
With this in hands it is easy to show that
\begin{equation}
\mH_{MN}\hat{D}_\a Y^M \hat{D}_\b Y^N= G_{mn}\dt_\a Y^m \dt_\b Y^n,
\end{equation}
where we used the fact that $A_{\m}{}^M=0$ for the chosen embedding of the brane. 

Finally, substituting all this into the expression for the entropy \eqref{inv_entr} one obtains
\begin{equation}
\label{Einv_NS5}
\begin{aligned}
S_{NS5}&=\int_\S d^5 \s e^{-2\f}\sqrt{G}|\tk|\fr{1}{\sqrt{G}}\sqrt{-\displaystyle \Big(g_{\m\n}\dt_\a X^\m \dt_\b X^\n+G_{mn}\dt_\a Y^m \dt_\b Y^n\Big)}\\
&=|\tk|\int_\S d^5 \s e^{-2\f}\sqrt{-\displaystyle \Big(g_{\m\n}\dt_\a X^\m \dt_\b X^\n+G_{mn}\dt_\a Y^m \dt_\b Y^n\Big)},
\end{aligned}
\end{equation}
which is the conventional expression for the geometric entanglement entropy of Ryu and Takayanagi (for the chosen embedding, i.e. $g_{\m m}=0$).

The same calculation can be repeated for the KK5-monopole. One starts with the following Killing vectors
\begin{equation}
k_a{}^M=(0,k_4^m;\tk_{e m}),
\end{equation}
where $e,f,g,h=1,2,3$. And the direction $4$ is identified with the Taub-NUT direction (the special circle of the monopole). For further convenience it is natural to choose such basis for the vectors $k_a^M$ where $\tk_{e 4}=0$ and $k_4^i$=0. Then the matrix $h_{ab}=k_a{}^Mk_b{}^N\mH_{MN}$ becomes
\begin{equation}
\begin{aligned}
h_{ef}&=\tk_{e m}\tk_{fn}g^{mn}=\tk_{e i}\tk_{f j}G^{ij},\\
h_{e4}&=0,\\
h_{44}&=k_4^4k_4^4 G_{44},
\end{aligned}
\end{equation}
and $\det h_{ab}=|\tk|^2 (k_4^4)^2 g^{-1} G_{44}$, where $g=\det g_{ij}$ is determinant of the 3-dimensional part of the metric $G_{mn}$ defined as
\begin{equation}
\begin{aligned}
G_{ij}&=g_{ij}+A_iA_j G_{44}, && & G_{i4}&=A_i G_{44},\\
G^{ij}&=g^{ij}, && & G^{i4}&=-A_{i4}G_{44},\\
G_{44}&=H^{-1},  && & G^{44}&=\fr{1}{G_{44}}+A_i A_i G_{44}.
\end{aligned}
\end{equation}
Following the same procedure as before it is straightforward to obtain the following expression for derivatives of the fields $Y^M$
\begin{equation}
\begin{aligned}
\hdt_\a Y^i&= \dt_\a Y^i, && &\hdt_\a Y^4&=-A_i \dt_\a Y^i\\
\hdt_\a \tY_i&=A_i \dt_a \tY_4, && &\hdt \tY_4&=\dt_\a \tY_4.
\end{aligned}
\end{equation}
The crucial difference between NS5 brane and KK5-monopole here is that for the former one is left only with the fields $Y^i$, which upon embedding into the full DFT action are identified with the usual geometric coordinates. In contrast, for KK5-monopole after projection one has the fields $\{Y^i,\tY_4\}$ which results in dependence of the background fields on the corresponding dual (winding) coordinate $\tx_4$. This behaviour has been observed in \cite{Harvey:2005ab,Jensen:2011jna,Bakhmatov:2016kfn} for KK5 and in \cite{Kimura:2013zva} for the exotic $5_2^2$-brane.

Finally, collecting all these pieces together one arrives at the following expression for entanglement entropy of the world-volume theory on (localized) Kaluza-Klein monopole
\begin{equation}
\label{Einv_KK55}
\begin{aligned}
S_{KK5}&=|\tk||k_4^4|\int_\S d^5 \s e^{-2\f} G_{44}\times\\
&\sqrt{-\displaystyle \det\Big[g_{\m\n}\dt_\a X^\m \dt_\b X^\n + \big(G_{ij}-G_{44}A_iA_j\big)\dt_\a Y^i \dt_\b Y^j +\big(G^{44}-G^{ij}A_iA_j\big)\dt_\a \tY_4 \dt_\b \tY_4\Big]}\\
&=|\tk||k_4^4|\int_\S d^5 \s e^{-2\f}G_{44}\sqrt{-\displaystyle \det\Big[g_{\m\n}\dt_\a X^\m \dt_\b X^\n + H\big(\d_{ij}\dt_\a Y^i \dt_\b Y^j +\dt_\a \tY_4 \dt_\b \tY_4\big)\Big]}.
\end{aligned}
\end{equation}
Where the last line is obtained by substituting the explicit background of KK5-monopole inside the square root. Taking into account that for the monopole one has $e^{-2\f}=1$ and $G_{44}=H^{-1}$ the second line reproduces precisely the expression \eqref{Einv_NS5} up to replacement $Y^4 \to \tY_4$. Note however, that talking about world-volume dynamics and field theories on the branes one does not distinguish between fields $Y^m$ and their duals $\tY_m$. The only difference is that the latter see the background T-dual to the background seen by the former. This is a trivial consequence of the above considerations.

Now, for exotic branes $5_2^r$ with $r=2,3,4$ the story is precisely the same and the algorithm is the following: fix the Killing vectors as in \eqref{Kill}, calculate $h_{ab}$ and hatted derivatives $\hdt_\a$, substitute everything in \eqref{inv_entr}. The result will always be \eqref{Einv_NS5} with the corresponding replacement of the fields $Y^m$ by their duals. Hence the name ``invariant entropy''. We postpone speculations on the physical and geometrical meaning of this procedure to the next section.

\section{Discussion}

In this letter we propose a T-duality invariant generalization of the Ryu-Takayanagi formula for geometric entanglement entropy for the case of NS  five-branes $5_2^r$ ($r=0,\dots,4$), with tension proportional to $g_s^{-2}$. The result is the expression \eqref{inv_entr} which is based on the same ideas as the effective action \eqref{full_5} for these branes. In particular, to choose a representative brane from the orbit one must specify Killing vectors, which satisfy the so-called algebraic section constraint. The choice which gives the effective action of NS5B-brane also reproduces the RT-formula for entanglement entropy of $\mc{N}=(1,1)$ Little String Theory living on this brane.

We check, that the same expression gives always the same result irrespective of which representative is chosen. This is in consistency with the fact, that e.g. the world-volume theory for the KK5A-brane is also $\mc{N}=(1,1)$ LST and hence the entropy should be the same. As we show in Section \ref{direct_KK} this is in contrast with the direct application of the Ryu-Takayanagi formula, which gives different results. 

On the level of world-volume scalar fields $Y^M$ transition between orbit representatives (say NS5B and KK5A) is just replacement of a field $Y^m$ by its duality partner $\tY_m$. Although this has crucial impact on DFT and supergravity solutions changing the background, the world-volume theory has no way to see that, and hence it is always the same. For this reason, as the carrier of the $\mc{N}=(1,1)$ LST in Type IIA string theory one should consider the localized Kaluza-Klein monopole rather than the smeared one \cite{Harvey:2005ab}. The former is a deformation of the latter by instanton corrections, and is already exotic since its harmonic function depends on one dual coordinate \cite{Jensen:2011jna,Bakhmatov:2016kfn}. The same is true for other exotic branes, which should also be localized.

The apparent issue that needs clarification is the following. For a theory on a Dp-brane one has apparent geometric picture, where the theory lives on a timelike surface at some $r\neq 0$ in the transverse space. For AdS/CFT correspondence one literally takes the conformal boundary of the anti-de-Sitter space. To calculate entanglement entropy geometrically one chooses a region $A$ on this surface and a surface $\S$ in the transverse space of the brane such that $\dt \S =\dt A$, and calculates its area in the given background. 

In the case in question one cannot develop such simple geometric picture, and moreover one cannot do this already for the NS5 brane. Indeed, the corresponding geometry does not drop into AdS and the corresponding field theory is not conformal. However, the $6D$ field theory associated with the brane can be as well put at any $\r$ in the transverse space, and the choice corresponds to the RG flow and one can still calculate area properly. The procedure described here suggests the following:
\begin{itemize}
\item start with a $5_2^r$-brane with any $r\in\{0,1,2,3,4\}$ and its world-volume theory described by the doubled amount of scalar fields $\F^M=(\F^m,\tilde{\F}_m)$ half of which is projected out by the algebraic section constraint;
\item choose a region $A$ in the space of the theory with boundary $\dt A$;
\item consider a surface $\S$ with boundary $\dt\S$ parametrized by some coordinates $\s^\a$;
\item this surface carries a doubled amount of scalar fields $\{Y^M\}$ with boundary conditions $Y^M\big|_{\dt\S}=\F^a\big|_{\dt A}$;
\item minimize the functional \eqref{inv_entr}.
\end{itemize}
The theory in the first item here just descents from the full invariant effective action \eqref{full_5}. The boundary condition is needed to identify the scalar fields living on the artificial surface $\S$ with the actual fields of the theory. For the conventional geometric picture this is done automatically by the embedding functions, where both the theory and the surface live in a single geometric background. Apparently, this procedure trivially reproduces the conventional geometric calculation, and the only messages here are the following:
\begin{itemize}
\item to calculate entanglement entropy for the $\mc{N}=(1,1)$ and $\mc{N}=(2,0)$ $6D$ theories one may use equivalently any of the representative of the T-duality orbit;
\item to get the correct result one must take into account proper localization of the backgrounds in the dual space.
\end{itemize}

An interesting  further direction of research is to generalize the expression to the case of M5-brane which belongs to the same orbit as the $5^3$-brane under U-duality group. One then still works with Little String Theory and 6D, however the invariant expression will be different. One can also consider D-branes in DFT, which can also be non-geometric, i.e. localized in the dual space. The corresponding effective action will be presented in the forthcoming paper \cite{axel-eric-fabio} and investigation of the corresponding world-volume theories and their entanglement entropy we reserve for future work.

\section*{Acknowledgements}  The authors are grateful to the Istanbul center of mathematical sciences and Bogazici University for hospitality during initial stages of this project. ETM would like to thank for hospitality Bogolyubov laboratory,  JINR, Dubna. The work of ETM was supported by the Russian state grant Goszadanie 3.9904.2017/8.9 and by the Alexander von Humboldt return fellowship and partially by the program of competitive growth of Kazan Federal University. 

\bibliography{bib}
\bibliographystyle{utphys}

\end{document}